%% file: reviewNR.tex
\documentclass[letter,prd,aps,twocolumn,floatfix,superscriptaddress,nofootinbib]{revtex4}
\usepackage{lipsum}
\usepackage{amssymb,amsmath,graphicx}
\usepackage{hyperref}
\usepackage[usenames,dvipsnames]{color}
\usepackage{array}
\usepackage{footnote}

\makeatletter
\renewcommand*{\p@section}{\S\,}
\renewcommand*{\p@subsection}{\S\,}

\makeatother

\input{aas_macros.tex}

\begin{document}

\title[Introduction to Numerical Relativity]{Introduction
	to Numerical Relativity} 

\author{Carlos Palenzuela}
\affiliation{Departament  de  F\'{\i}sica $\&$ IAC3,  Universitat  de  les  Illes  Balears,  Palma  de Mallorca,  Baleares  E-07122,  Spain}

\begin{abstract}
	
	Numerical Relativity is a multidisciplinary field including relativity, magneto-hydrodynamics, astrophysics and computational methods, among others, 
	with the aim of solving numerically highly-dynamical, strong-gravity scenarios where no other approximations are available. Here we describe some of the foundations of the field, starting from the covariant Einstein equations and how to write them as a well-posed system of evolution equations, discussing the different formalisms, coordinate conditions and numerical methods commonly employed nowadays for the modeling of gravitational wave sources. 
	
\end{abstract}

\maketitle

\section{Introduction}

General Relavity is the theory that identifies gravity as the curvature of a four dimensional space-time manifold.
The consequences of this identification deeply changed our conception of Nature. From the physics point of view, Relativity introduced new ideas, like that time and space are not absolute but depend on the observer, that the effects of gravity propagate at the speed of light, or that energy and matter are equivalent and can modify the structure of both space and time, among others.
From the mathematical point of view, the main consequence is that gravity can be described by using the tools of differential geometry, where the basic object to represent a manifold is the metric $g_{ab}$ that allow us to compute distances between neighboring points. The famous Einstein equations describe the dynamics of the four-dimensional space-time metric and how it is deformed by a given mass-energy distribution. On the other hand, the Bianchi identities from differential geometry ensure that the divergence of the Einstein tensor vanishes, implying the conservation of the stress-energy tensor (i.e., corresponding to energy and linear momentum conservation), which describes how matter moves in a curved spacetime.

One of the greatest achievements of General Relativity was the prediction of gravitational waves, space-time deformations produced by acceleration of masses which behave like waves as they propagate away from the sources. Gravitational waves are essentially unscattered between emission and detection, thereby giving direct information about the sources powering these phenomena. Precisely due to the weak interaction of these waves with matter, their existence was initially only confirmed indirectly by observations of the orbital dynamics of binary pulsars~\cite{1975ApJ...195L..51H,2014LRR....17....4W}. However, current kilometer-scale interferometric gravitational wave (GW) detectors, such as Advanced LIGO (aLIGO)~\cite{Abramovici325} and Advanced Virgo (adVirgo)~\cite{1997CQGra..14.1461C} facilities, since 2015 have directly detected gravitational waves on the kiloHertz frequency regime, consistent with the merger of binary black holes and binary neutron stars~\cite{2019PhRvX...9c1040A}. Further improvements on these detectors, as well as new ones added to the array of GW observatories, will allow to establish many routinary GW observations in the next few years. These new observations allow us a new way to study some of the most energetic and exotic processes in the universe and start a new era of gravitational wave astronomy that will inevitably lead to unprecedented discoveries and breakthroughs not only in Astrophysics and Cosmology, but also in fundamental theories like gravity and nuclear physics. The detection, identification, and accurate determination of the physical parameters of sources is crucial to validate (and challenge) not only our theories but also our astrophysical models, which rely both on precise experimental data and on the availability of template banks of theoretical waveforms. For the slow inspiral, when the neutron stars (NSs) or black holes (BHs) are widely separated, analytical approximations for the gravitational waveforms are provided by perturbative post-Newtonian (PN) expansion techniques~\cite{2014LRR....17....2B}. For the last orbits and merger, where the fields are particularly strong and most might be gained in terms of insight on fundamental physics, the PN expansion breaks down and the full Einstein equations have to be solved numerically. This has only become possible after a series of breakthroughs in the field of Numerical Relativity~\cite{PhysRevLett.95.121101,PhysRevLett.96.111101,PhysRevLett.96.111102}, calling for an incorporation of this new type of information into data analysis strategies and methods. Since then, outstanding progress has been made to explore the late stage of binary coalescence with numerical methods. The next sections summarize some of the foundations of
Numerical Relativity,  with a view on the modeling of gravitational sources, from the construction of a well-posed evolution system to the numerical methods commonly employed to solve them. Notice that this review focus on Cauchy formulations, excluding other alternatives. For a wider overview of all the possible formulations, please see ~\cite{2001CQGra..18R..25L}.

\section{Evolution systems}

\subsection{Einstein Equations}

The equations of motion of a classical theory like General Relativity can be derived directly from a suitable action by using the Euler-Lagrange equations, leading to the well-known Einstein equations~\cite{Misner_1973},
\begin{eqnarray}\label{EE1}
G_{ab} \equiv R_{ab} - \frac{R}{2} g_{ab} = 8 \pi  T_{ab} ~~~~,~~
\end{eqnarray}
where $G_{ab}$ is the Einstein tensor, $R_{ab}$ is the Ricci tensor of the spacetime represented by the metric $g_{ab}$, $R \equiv g^{ab} R_{ab}$ is the Ricci or curvature scalar, and $T_{ab}$ is the  stress-energy tensor describing generically the matter-energy distributions in the spacetime. We have chosen geometric units such that $G=c=1$ and adopt the convention where roman indices $a,b,c,...$ denote space-time components  (i.e.,  from  0  to  3),  while $i,j,k,...$ denote  spatial ones (i.e., from 1 to 3). 

The Ricci tensor can be written in terms
of the Christophel symbols $\Gamma^a_{bc}$ as follows
\begin{eqnarray}\label{EE2}
R_{ab} &=& \partial_c \Gamma^{c}_{ab} -
\partial_a \Gamma^{c}_{cb} + 
\Gamma^{c}_{cd} \Gamma^{d}_{ab}
-\Gamma^{c}_{da} \Gamma^{d}_{cb}
~~~~,~~~~ \\
\Gamma^{c}_{ab} &=& \frac{1}{2} g^{cd} \left(
\partial_a g_{bd} + \partial_b g_{ad} - \partial_d g_{ab} \right) ~~,
\label{EE2b}
\end{eqnarray}
Notice that Eqs.~(\ref{EE1}-\ref{EE2b}) form a system of ten non-linear partial differential equations (PDEs) for the spacetime metric components $g_{ab}$, which are coupled to the matter fields by means of the stress-energy tensor.

On the other hand, an important relation in differential geometry, known as the (contracted) Bianchi identities, implies the covariant conservation law for the Einstein tensor and, consequently, for the stress-energy tensor, 
\begin{eqnarray}\label{EE3}
\nabla_a G^{ab} &=& 0 \Longrightarrow \nabla_a T^{ab} = 0  ~~~~~,~~~~~ 
\\
 \nabla_a T^{ab} &=& \partial_a T^{ab} + \Gamma^{a}_{ac} T^{cb} + \Gamma^{b}_{ac} T^{ac} 
\end{eqnarray}
where $\nabla_a$ is the covariant derivative, the generalization of the partial derivative on a manifold.
These covariant equations correspond to conservation laws for both the energy and linear momentum, which are the basic physical equations to describe any matter field.
Notice also that the Bianchi identities imply that four of the ten components of Einstein's equations cannot be independent. This redundancy gives rise to both the four coordinate degrees freedom and the four constraint equations, which will be clearly manifested in the 3+1 decomposition described in the next section.

\subsection{The 3+1 decomposition}

Despite its elegance and compactness, the covariant form of the four-dimensional Einstein equations is not suitable to describe how the gravity fields evolve from an initial configuration towards the future. In such case, it is more intuitive to  consider instead a time succession of three-dimensional spatial slice
geometries, called foliation, where the evolution of a given slice is given by the
Einstein equations (for more detailed treatments see for instance~\cite{2007gr.qc.....3035G,2008itnr.book.....A, 2009LNP...783.....B, 2010nure.book.....B,10.5555/2904075}).
This 3+1 decomposition, in which the four-dimensional manifold is splitted into ``space+time'' components and the 
covariant Einstein equations are converted into evolution equations for three-dimensional geometric fields, can be summarized in the following steps:
\begin{itemize}
	\item \emph{specify the  choice of coordinates. }
	The covariance of Einstein equations implies that they can be written in the same generic way on any system of coordinates, which can be defined by a set of observers. The spacetime can be foliated by a family of spacelike hypersurfaces $\Sigma$, which are intersected by a congruence of time lines that will determine
	our observers (i.e., our system of coordinates). This congruence is described by
	the vector field $t^a = \alpha n^a +\beta^a$, where
	$n^a$ is the timelike unit vector normal to the spacelike hypersurfaces, $\alpha$ is the
	lapse function which measures the proper time of the Eulerian (orthogonal) observers and
	$\beta^a$ is the shift vector that measures the displacement, between consecutive hypersurfaces, of the time line $t^a$ followed by the observers with respect to the normal $n^a$ (see Figure~\ref{3+1decomposition}).

	\begin{figure}
		\centering
		\includegraphics[width=0.8\linewidth]{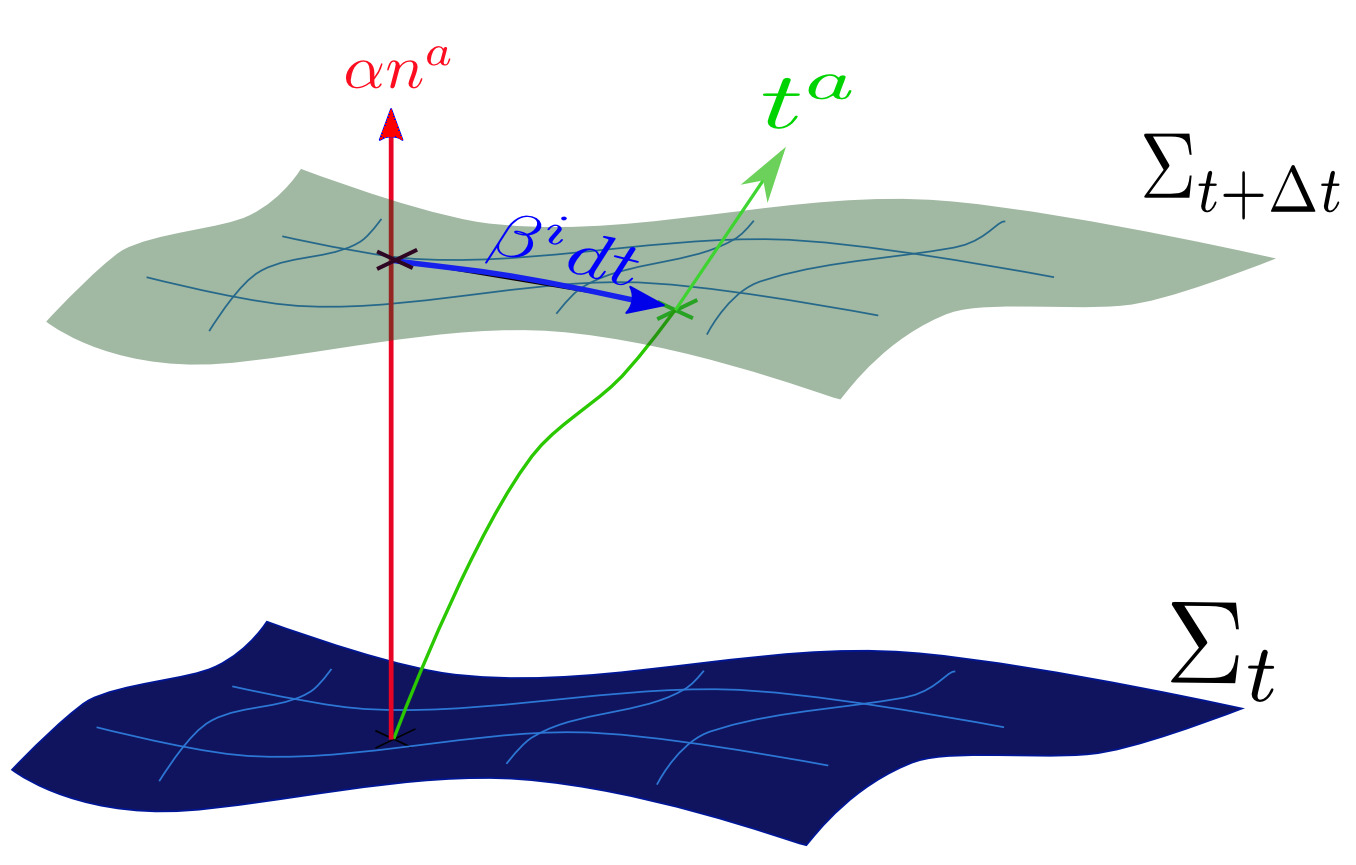}
		\caption{{\em Foliation of the spacetime manifold.} The lapse function $\alpha$ measures the proper time along the normal $n^a$ to the hypersurface $\Sigma_t$, which is equipped with an induced metric $\gamma_{ij}$. The shift vector $\beta^{i}$ measures the displacement, on consecutive hypersurfaces, between the observer time lines $t^a$ and the normal lines $n^a$.}
		\label{3+1decomposition}
	\end{figure}

	\item \emph{decompose every 4D object into its 3+1 components.}
	The choice of coordinates allows for the definition of a spatial projection tensor ${\gamma^a}_b \equiv \delta^a_b + n^a\, n_b$. Any four-dimensional tensor can
	be decomposed into 3+1 pieces using either the spatial projector to obtain
	the spatial components, or contracting with $n^a$ for the time
	components. For instance, the line element measuring the distance between neighboring points can be written by using these generic 3+1 coordinates as
	\begin{eqnarray}\label{line_element}
	ds^2 &=& g_{ab} dx^a dx^b \nonumber \\
	&=& - \alpha^2\, dt^2 + \gamma_{ij} (dx^i + \beta^i dt)\,
	(dx^j + \beta^j dt) ~~~,
	\end{eqnarray}
	where the spatial three-dimensional induced metric $\gamma_{ij}$ is just the projection of the four-dimensional metric $g_{ab}$ into the space-like hypersurface $\Sigma$. Other objects, like the stress-energy tensor, can also be decomposed into its various components, namely
	\begin{equation}\label{3+1_stress_energy}
	\tau \equiv T^{ab}\, n_a\, n_b ,~ S_i \equiv - T_{ab}\,  n^a \,{\gamma^b}_i ,~
	S_{ij} \equiv T_{ab}\, {\gamma^a}_i\, {\gamma^b}_j 
	\end{equation}
	
	\item \emph{write down the field equations in terms of the 3+1 components.}
	Within the framework outlined here, the induced 
	metric $\gamma_{ij}$ is the only unknown, since both lapse and shift are set by our choice of coordinates.
	In differential geometry it is also common to define
	an additional tensor $K_{ij}$ with a strong geometrical meaning, as it describes the change of
	the induced metric along the congruence of normal observers. This definition involves the Lie derivative ${\mathcal L}_{{\bf n}}$, a generalization of the directional derivative along the vector ${\bf n}$ in a manifold. Therefore, the definition of the extrinsic curvature and the 3+1 decomposition of Einstein equations form an hyperbolic-elliptic system of PDEs, commonly known as the Arnowitt-Deser-Misner (ADM) formalism~\cite{1962gait.book.....W,york79}, which can be written as	
	\begin{eqnarray}\label{Kdef}
	K_{ij} &\equiv& -\frac{1}{2}
	{\mathcal L}_{{\bf n}} \gamma_{ij} = - \frac{1}{2\alpha} \left( \partial_t-{\mathcal L}_\beta \right)\gamma_{ij} ~~~,~~~ \\
	{\mathcal L}_\beta \gamma_{ij} &=& \beta^k \partial_k \gamma_{ij} + \gamma_{ik} \partial_j \beta^k
	+ \gamma_{kj} \partial_i \beta^k ~,
	\end{eqnarray}	
	\begin{eqnarray}
	\label{Kevol}
	\left(\partial_t - \mathcal{L}_\beta\right) K_{ij}
	&=& - \nabla_i  \nabla_j   \alpha
	+ \alpha \left(  R_{ij} -2{K_i}^k {K_{jk}} + {\rm trK}\,K_{ij} \right) \nonumber \\
	&-&  8\pi \alpha \left[S_{ij}-{\frac{\gamma_{ij}}{2}}\left({\rm trS} - \tau\right)\right]  ~,
	\end{eqnarray}
	\begin{eqnarray}
	\label{energyconst}
	\mathcal{H} &=& {R_i}^i + \left({\rm trK}\right)^2 -  {K_i}^j\, {K_j}^i - 16\, \pi\, \tau = 0 ~, \\
	\label{momconst}
	\mathcal{M}_i &=& \nabla_j\;\left({K_i}^j - {\rm trK}\;{\delta_i}^j \right) - 8\, \pi\, S_i = 0 ~.
	\end{eqnarray}
\end{itemize}
where we have defined the trace of any three-dimensional tensor $C_{ij}$ as $tr C = \gamma^{ij} C_{ij}$. The evolution hyperbolic equations (\ref{Kdef},\ref{Kevol}) for the evolved fields $\{ \gamma_{ij}, K_{ij} \}$
are complemented with the energy and momentum constraint equations (\ref{energyconst},\ref{momconst}), that have to be satisfied at each hypersurface. This system of equations  needs to be completed with a specification of the coordinate system, that is, by a choice of lapse and  shift $\{ \alpha, \beta^i\}$. The ADM formalism still preserves the covariance under spatial or time coordinate transformations (i.e., 3+1 covariance). Notice that, although manifest four-dimensional covariance is lost when performing the 3+1 decomposition, the solution space is still invariant under general coordinate transformations.

One can take advantage of the contracted Bianchi identities to prove that the constraint equations (\ref{energyconst},\ref{momconst}) are just first integrals of the evolution ones (\ref{Kdef},\ref{Kevol}), so that if the constraints are satisfied on an initial hypersurface (i.e., $\mathcal{H}=\mathcal{M}_i=0$ at $\Sigma_t$), they will remain satisfied for all times. This redundancy of the equations allows for different evolution approaches. 
The most straightforward choice, the {\it constrained evolution} approach, involves solving simultaneously both the evolution equations and the constraints, but it presents several difficulties. From the theoretical point of view, it is not clear how to split the dynamical modes, solved through evolution equations, from the constrained ones, enforced by elliptic equations. From the computational point of view, elliptic equations are computationally more expensive and difficult to solve efficiently than hyperbolic ones. A simpler alternative is given by the {\it free evolution} approach, where the fields are obtained uniquely from the evolution equations, while the constraints are enforced only at the initial time (i.e., although they can be computed during the evolution to estimate the validity of the solution). Notice however that discarding the constraint equations breaks the underlying invariance of the solutions.
Due to its simplicity, the free evolution approach has traditionally been the most common choice in Numerical Relativity applications without symmetry assumptions,
particularly in efforts associated to the modeling of gravitational wave sources.

It is important to stress that any astrophysical scenario, except those including only black holes, involves some type of matter, which will evolve on a curved spacetime as described by Eq.~\ref{EE3}. We can also perform the 3+1 decomposition on this equation to obtain the evolution for the matter energy density $\tau$ and the momentum density $S_i$, namely
\begin{eqnarray}
(\partial_t - \mathcal{L}_{\beta}) \tau 
&+& \alpha \nabla_k\, S^k \nonumber \\
 &=& \alpha \left(\tau\, trK - 2 S^k \, \partial_k \ln \alpha +  K_{ij} S^{ij} \right) ~~,
\label{fluid31a} \\
(\partial_t - \mathcal{L}_{\beta}) S_i
&+& \alpha \nabla_k \, {S^k}_i \nonumber \\
&=& \alpha \left(S_i\, trK - {S^k}_{i} \, \partial_k \ln \alpha - \tau\, \partial_i \ln \alpha \right) ~~.
\label{fluid31b}
\end{eqnarray}
Notice that these equations need a closure relation   ${S^k}_i = {S^k}_i (\tau,S_i)$ that will depend on the type of matter considered. 

\subsection{Formulations of the Einstein Equations}

Any mathematical model representing a physical system must be described by a well-posed system of equations, meaning that there exists a unique bounded solution that depends continuously on the initial data. Such requirement is relevant not only from
a conceptual point of view, but it is of crucial importance in computational applications: a numerical solution solving an ill-posed problem is not enforced to converge to its corresponding continuum solution.
A clear example of this undesired behavior can be observed in the ADM free evolution system resulting directly from the 3+1 decomposition of Einstein equations. Although
the ADM formalism was extensively used at the dawn of Numerical Relativity due to its simplicity, the presence of several numerical instabilities in the three-dimensional case made it unsuitable for computational applications. The  reason behind these instabilities, as it was shown in the nineties,  was the ill posedness of the ADM system in 3+1 dimensions when supplemented with standard gauge conditions. 

Since then, there have been several attempts to construct well-posed free-evolution formalisms, either by selecting a particular gauge or by mixing the constraints with the evolution equations to modify the principal part of the system. The mathematical structure of the Einstein field equations was first investigated on a specific coordinate choice, called the harmonic gauge,
in which the spacetime coordinates follow wave equations and can be written as $\Gamma^a \equiv g^{bc} \Gamma^{a}_{bc} = 0$ \cite{1962gait.book.....W}.
This choice allowed to greatly simplify Einstein equations, which could then be written as a set of (well-posed) generalized wave equations,  $g^{cd} \partial_c \partial_d g_{ab} = H_{ab}(g, \partial g)$, where $H_{ab}$ is a quadratic function in the metric first derivatives. 
This Harmonic formalism, written for different set of fields and for generalized harmonic conditions~\cite{2005CQGra..22..425P,2006CQGra..23S.447L}, was used successfully to model the coalescence of compact objects, like black holes~\cite{2006CQGra..23S.529P,2009PhRvD..80l4010S}, boson stars~\cite{2007PhRvD..75f4005P} and neutron stars~\cite{Anderson_2008}.

Another very convenient way to write down Einstein equations is the Baumgarte-Shapiro-Shibata-Nakamura(BSSN)  formalism\cite{1995PhRvD..52.5428S,1999PhRvD..59b4007B}, which relies in three important modifications of the ADM system. First, it applies a conformal decomposition on the evolved fields, partially motivated by the fact that the Schwarschild black hole solution is conformally flat. Therefore, a conformal metric ${\tilde \gamma}_{ij}$, with unit determinant, and a conformal, trace-less, extrinsic curvature ${\tilde A}_{ij}$ can be introduced as 
\begin{eqnarray}
{\tilde \gamma}_{ij} = \chi \,\gamma_{ij} ~~~~,~~~~ 
{\tilde A}_{ij} = \chi \, \bigl( K_{ij} - {1\over3} \, \gamma_{ij} \, tr K \bigr) ~.
\end{eqnarray} 
These new definitions involve the appearance of two new constraints, ${\tilde \gamma} = 1$ and $
{tr \tilde A} \equiv {\tilde \gamma}^{ij} {\tilde A}_{ij}  = 0$, which will be denoted as {\em conformal constraints} from now on
to distinguish them from the energy-momentum {\em physical constraints}. The second modification consists on extending the space of solutions by introducing a new evolved field  ${\tilde \Gamma}^i = {\tilde \gamma}^{jk} \, {\tilde \Gamma}^i{}_{jk} = - \partial_j {\tilde \gamma}^{ij}$, namely the contraction of the Christoffel symbols associated to the conformal metric. The third modification, which is essential to achieve a well-posed system, is to add the momentum constraint in a specific way to the evolution equation for this new quantity ${\tilde \Gamma}^i$ (i.e., which is originally calculated, as usual, by taking the time derivative of its definition). Notice that the last two modifications are analogous to
rewrite the momentum constraint as an evolution equation and affect strongly the principal part of the system (i.e., the terms with derivatives of highest order), transforming the free-evolution ADM ill-posed system into a well-posed one, when supplemented with appropriate gauge conditions~\cite{2002PhRvD..66f4002S,2006PhRvD..74b4016G}.
This formalism, with the 1+log slicing and the gamma-freezing shift conditions described below, has been used successfully to model the coalescence of black holes without the need of excising the interior of the apparent horizons to remove the physical singularity from the computational domain~\cite{2006PhRvD..73l4011V,2008PhRvD..77b4027B}, making them especially convenient for black hole simulations~\cite{2006PhRvD..73j4002B,2006PhRvD..73f1501C,2007PhRvD..76j4015S}. Notice however that the BSSN formalism was already being used successfully to model the coalescence of binary neutron stars~\cite{2000PhRvD..61f4001S,2002PThPh.107..265S}, although the lack of advanced computational techniques like Adaptive Mesh Refinement(AMR) prevented the calculation of accurate waveforms until several years later.

An asymmetry of the BSSN formalism is manifested on the different ways to treat the physical constraints, since the momentum constraint is mixed with the evolution equations but the energy constraint is not. Related to this, and like many other contemporary formalisms, BSSN does not include any mechanism to control dynamically unavoidable constraint violations, which could grow significantly during a numerical simulation, even if they are only seeded by tiny discretization errors~\cite{2002PhRvD..66h4014L}. The Z4 formalism, which was introduced as a  extension of the Einstein equations to achieve a well posed, hyperbolic evolution system free of constraints~\cite{2003PhRvD..67j4005B,2004PhRvD..69f4036B}, allowed to address these issues in an elegant general-covariant way.
The equations of motion can be derived from a suitable action via a Palatini-type variation~\cite{2010PhRvD..82l4010B}, obtaining
\begin{eqnarray}\label{Z4cov}
R_{ab} + \nabla_a Z_b + \nabla_b Z_a   &=& 
8\pi \, \left( T_{ab} - \frac{1}{2}g_{ab} \,tr T \right) \\
&+& \kappa_{z} \, \left(  n_a Z_b + n_b Z_a - g_{ab} n^c Z_c \right) ~,
\nonumber 
\end{eqnarray}
where $Z_{a}$ is introduced as a new four-vector measuring the deviation from Einstein's solutions, which are those satisfying the algebraic condition $Z_{a}=0$. Although the original formulation, corresponding to the choice $\kappa_{z}=0$, is completely covariant, additional damping terms were included to enforce dynamically the decay of the physical constraint violations associated to $Z_a$~\cite{1999JMP....40..909B}. As it is shown in~\cite{Gundlach:2005eh}, all the physical constraint modes are exponentially damped if $\kappa_{z} >0$. However, since the damping terms are proportional to the unit normal of the time slicing $n_{a}$, the full covariance of the system is lost due to the presence of this privileged time vector.
The 3+1 decomposition of the Z4 formalism given by eq.~(\ref{Z4cov}) leads to evolution equations for the evolved fields $\{ \gamma_{ij}, K_{ij}, Z_i, \Theta  \}$, where we have defined the normal projection $\Theta \equiv - n_{a} Z^{a}$. Notice that now there are ten evolution equations to solve ten unknowns; the original elliptic constraints in the Einstein Equations have been converted into evolution equations for the new four-vector $Z_a$, which can be understood roughly as the time integral of the energy and momentum constraints. Einstein's solutions are recovered when the algebraic constraint $Z_a=0$ is satisfied. Finally, the most important feature is that the evolution system, when combined with suitable gauge conditions, is directly well-posed, without the need of further modifications~\cite{2004PhRvD..69j4003B}.

The Z4 formalism has also been useful to understand also the constraint evolution system (i.e., subsidiary system) and the connection among different formalisms. For instance, the Harmonic formalism can be recovered from the Z4 one by substituting the harmonic condition with $\Gamma^a = - 2 Z^a$~\cite{2003PhRvD..67j4005B}, and a version of the BSSN by a symmetry-breaking mechanism~\cite{2004PhRvD..69f4036B}. 
Along these lines, one can take advantage of the Z4 formalism flexibility to incorporate the ability to deal with black hole singularities without excision. The conformal and covariant $Z4$ (CCZ4) formalism~\cite{2012PhRvD..85f4040A} was constructed by performing the same conformal transformations as in the BSSN formalism (i.e., see also~\cite{2010PhRvD..81h4003B} for other conformal but non-covariant Z4 formulations) but using, instead of $trK$ and $Z_i$, the following quantities as evolved fields,
\begin{eqnarray}
{tr \hat K} \equiv tr K - 2\, \Theta ~~~~~,~~~~~ 
{\hat \Gamma}^i \equiv {\tilde \Gamma}^i + {2 \over \chi}  Z^{i} ~, 
\end{eqnarray}
so that the evolution equations are closer to those in the BSSN formulation. The full list of evolved fields is then given by
$\{ \chi, {\tilde \gamma}_{ij}, {tr \hat K}, {\tilde A}_{ij}, {\hat \Gamma}^i, \Theta  \}$ and follow the evolution equations~\cite{2017PhRvD..95l4005B}, 
\begin{eqnarray}
\partial_t {\tilde \gamma}_{ij} 
& =& \beta^k \partial_k {\tilde \gamma}_{ij} + {\tilde \gamma}_{ik} \, \partial_j \beta^k 
+ {\tilde \gamma}_{kj} \partial_i \beta^k - {2\over3} \, {\tilde \gamma}_{ij} \partial_k \beta^k \nonumber \\
&-& 2 \alpha \Bigl( {\tilde A}_{ij}  - \frac{1}{3} {\tilde \gamma}_{ij}\, tr {\tilde A} \Bigr) -  \frac{\kappa_{c}}{3}\,\alpha\tilde{\gamma}_{ij}\ln\tilde{\gamma}, \label{syseq1}
\\
\partial_t {\tilde A}_{ij} 
& =& \beta^k \partial_k{\tilde A}_{ij} + {\tilde A}_{ik} \partial_j \beta^k 
+ {\tilde A}_{kj} \partial_i \beta^k 
\nonumber \\
&-& {2\over3} \, {\tilde A}_{ij} \partial_k \beta^k - \,\frac{\kappa_{c}}{3}\,\alpha\,\tilde{\gamma}_{ij}
\,tr \tilde{A}
\nonumber \\
& +& \chi \, \Bigl[ \, \alpha \, \bigl( {^{(3)\!}{\hat R}}_{ij} + {\hat R}^\chi_{ij} 
- 8 \pi \, S_{ij} \bigr)  - \nabla_i \nabla_j \alpha \, \Bigr]^{\rm TF} 
\nonumber \\
&+& \alpha \, \Bigl( tr {\hat K} \, {\tilde A}_{ij} - 2 {\tilde A}_{ik} {\tilde A}^k{}_j \Bigr), 
\\
\partial_t \chi & =& \beta^k \partial_k \chi 
+ {2\over 3} \, \chi \, \bigl[ \alpha (tr {\hat K} + 2\, \Theta) - \partial_k \beta^k  \bigr], 
\\
\partial_t tr {\hat K} 
& =&  \beta^k \partial_k tr {\hat K} 
- \nabla_i \nabla^i \alpha
+ \alpha \Big[ {1 \over 3} \bigl( tr {\hat K} + 2 \Theta \bigr)^2 
 \\
&+& {\tilde A}_{ij} {\tilde A}^{ij} + 4\pi  \bigl(\tau + tr S\bigr)
+  \kappa_z  \Theta \Big]  
+ 2\, Z^i \nabla_i \alpha, 
\nonumber \\
\partial_t \Theta 
& =&  \beta^k \partial_k \Theta + {\alpha \over 2} \left[ {^{(3)\!}R} + 2 \nabla_i Z^i
+ {2\over3} \, tr^2{\hat K} \right.
\nonumber \\
&+&  \left. {2\over3} \, \Theta \Bigl( tr {\hat K} - 2 \Theta \Bigr)
- {\tilde A}_{ij} {\tilde A}^{ij}  \right] - Z^i \nabla_i \alpha 
\nonumber \\
&-& \alpha \, \Bigl[ 8\pi  \, \tau  + 2\kappa_z  \, \Theta \Bigr], 
\\   
\partial_t {\hat \Gamma}^i 
& =& \beta^j \partial_j {\hat \Gamma}^i - {\hat \Gamma}^j \partial_j \beta^i 
+ {2\over3} {\hat \Gamma}^i \partial_j \beta^j + {\tilde \gamma}^{jk} \partial_j \partial_k \beta^i
\nonumber \\
&+& {1\over3} \, {\tilde \gamma}^{ij} \partial_j \partial_k \beta^k 
 - 2 {\tilde A}^{ij} \partial_j \alpha 
 + 2\alpha \, \Bigl[ {\tilde \Gamma}^i{}_{jk} {\tilde A}^{jk}
\nonumber \\ 
&-& {3 \over 2 \chi} \, {\tilde A}^{ij} \partial_j \chi 
- {2\over3} \, {\tilde \gamma}^{ij} \partial_j tr{\hat K} - 8\pi \, {\tilde \gamma}^{ij} \, S_i \Bigr] \nonumber
\\ 
& +& 2 \alpha \, \left[- {\tilde \gamma}^{ij} \left( {1 \over 3}\partial_j \Theta 
+ {\Theta \over \alpha} \, \partial_j \alpha \right) \right.
\nonumber \\
&-& \left. {1 \over \chi} Z^i \left( \kappa_z + {2\over 3} \, (tr{\hat K} + 2 \Theta) \right) \right],   
\label{syseq2}
\end{eqnarray}
where the expression $[\ldots]^{\rm TF}$ indicates the trace-free part with respect to the metric $\tilde{\gamma}_{ij}$. 
The non-trivial terms inside this expression can be written as 
\begin{eqnarray}
{\hat R}^{\chi}_{ij} & = & {1 \over 2 \chi} \, \partial_i \partial_j \chi 
- {1 \over 2 \chi} \, {{\tilde \Gamma}^k}_{ij} \partial_k \chi 
\nonumber \\
&-& {1 \over 4 \chi^2} \, \partial_i \chi \partial_j \chi 
+ {2 \over \chi^2} Z^k {\tilde \gamma}_{k(i} \partial_{j)} \chi 
\nonumber\\
&+& {1 \over 2 \chi }{\tilde \gamma}_{ij} \, \Bigl[ {\tilde \gamma}^{km} \Bigl( {\partial}_k {\partial}_m \chi 
-  {3\over 2 \chi} \, \partial_k \chi \partial_m \chi \Bigr)
- {\hat \Gamma}^k \partial_k \chi \Bigr] ~,
\nonumber     \\
{^{(3)\!}{\hat R}}_{ij} &=& - {1\over2} \, {\tilde \gamma}^{mn} \partial_m \partial_n {\tilde \gamma}_{ij}     
+ {\tilde \gamma}_{k(i} \partial_{j)} {\hat \Gamma}^k     
+  {\hat \Gamma }^k {\tilde \Gamma}_{(ij)k} 
\nonumber   \\
&+& {\tilde \gamma}^{mn} 
\Bigl(  {{\tilde \Gamma}^k}_{mi} {\tilde \Gamma}_{jkn} 
\Bigr.
+  {{\tilde \Gamma}^k}_{mj} {\tilde \Gamma}_{ikn} +{\tilde \Gamma}^k{}_{mi} {\tilde \Gamma}_{knj}  \Bigr)~, 
\nonumber \\
\nabla_i \nabla_j \alpha & = & \partial_i \partial_j \alpha - {{\tilde \Gamma}^k}_{ij} \partial_k \alpha 
\nonumber \\
&+& {1 \over 2 \chi} \Bigl( \partial_i \alpha\, \partial_j \chi 
+  \partial_j \alpha  \, \partial_i \chi
- {\tilde \gamma}_{ij}\, {\tilde \gamma}^{km}\, \partial_k \alpha\, \partial_m \chi  \Bigr) ~,  
\nonumber
\end{eqnarray}

Notice that damping terms proportional to a free parameter $\kappa_c$ have been included in order to dynamically control the conformal constraints, exactly in the same way as it is done with the physical ones.

\subsection{Gauge conditions}

The principle of general covariance implies that the laws of physics, and in particular Einstein equations, must take the same form for all observers. This implies that they have to be written in a generic tensor form for any system of coordinates. The choice of coordinates is commonly referred as gauge freedom, and it corresponds to define the congruence of our observers, i.e., the time vector $t^a$ by setting the lapse and shift.
Notice that setting gauge conditions is not only necessary to close the system of equations: these additional degrees of freedom can also be useful both to avoid coordinate or physical singularities and to adapt to the underlying symmetries appearing in our simulations. Besides the summary presented here, further details on the different gauge conditions can be found for instance in~\cite{2007gr.qc.....3035G,2008itnr.book.....A, 2009LNP...783.....B, 2010nure.book.....B,10.5555/2904075}.

The simplest gauge conditions, known as geodesic  coordinates, are obtained setting $\alpha=1$ and $\beta^{i}=0$, so that the time coordinate coincides with the proper timer of the Eulerian observers (i.e., those following timelike geodesics).
A simple perturbation analysis shows however that any formalism supplemented with this choice of coordinates might suffer of unstable non-physical modes. Even worse, this gauge condition might also lead to coordinate singularities, since Eulerian observers will focus into a single point such that the spatial volume $\sqrt{\gamma} \rightarrow 0$.
Coordinate pathologies can be prevented by imposing suitable geometrical conditions, which usually involve some type of elliptic equations~\cite{1978PhRvD..17.2529S}. This is the case, for instance, in the \textit{maximal slicing condition} $trK = 0$, which, when imposed at all times, implies 
\begin{equation}
\nabla^{i} \nabla_{i}\alpha = \alpha \left[ K_{ij}K^{ij} + 4\pi(\tau+S) \right]~.
\label{alphaellip}
\end{equation}
This slicing condition is called \textit{singularity-avoiding} condition because the lapse function $\alpha$ goes to zero when the spatial volume $\sqrt{\gamma}$ goes to zero, avoiding the coordinate singularities during
the evolution by slowing-down the proper time of the observers near strong-gravity regions. 
Another interesting geometrical property to be satisfied would be the \textit{minimal distortion condition}, which can be written as
\begin{equation}
\nabla^{j} \nabla_{j} \beta^i + \frac{1}{3} \nabla^i \nabla_j \beta^j + {R^i}_j \beta^j= 2 \nabla_j \left[\alpha (K^{ij} - \frac{1}{3} \gamma^{ij} trK)\right] ~~.
\label{betaellip}
\end{equation}
This shift condition minimizes the changes in the shape of the volume elements, independently of their size. Both the maximal slicing and the minimal distortion conditions~(\ref{alphaellip},\ref{betaellip}) are elliptic equations. These type of equations are computationally much more expensive than hyperbolic evolution ones, and are usually avoided or transformed into hyperbolic ones in the context of free evolution formalisms.

Indeed, hyperbolic evolution equations are preferred 
and were already adopted to enforce some interesting property, like for instance the harmonic coordinates, which ensured the well-posedness of the Harmonic formalism  \cite{1962gait.book.....W}. A suitable family of evolution equations for the lapse is given by the 
\textit{Bona-Mass\'o slicing condition}~\cite{1995PhRvL..75..600B}, 
\begin{eqnarray}
\partial_t \alpha & = &  \beta^i \partial_i \alpha - \alpha^{2} \,f(\alpha)\, trK~,
\label{alphacond}
\end{eqnarray}  
which, for any $f(\alpha) \geq 1$, is not only singularity avoiding, but also maintains the well-posedness of the formalism. The case $f(\alpha)=1$ correspond to the harmonic slicing condition, while that $f(\alpha)\to\infty$ mimics the maximal slicing condition Eq.(\ref{alphaellip}). A common choice in numerical applications, especially those involving black holes, is to use the so-called \textit{$1+\log$ slicing condition}, corresponding to $f(\alpha)=2/\alpha$. This choice has excellent singularity avoidance conditions, since near the physical singularity $\alpha \to 0$, mimicking the maximal slicing condition.

A suitable family of hyperbolic dynamical equations for the shift-vector $\beta^{i}$ is given by the \textit{Gamma-driver condition}~\cite{2003PhRvD..67h4023A}, 
\begin{eqnarray}
\partial_t \beta^i & =& \beta^j \partial_j \beta^i + \, g(\alpha) \, {\hat \Gamma}^i - \eta \beta^i  ~~,
\label{system3}
\end{eqnarray}
where $g(\alpha)$ is an arbitrary function depending on the lapse function and $\eta$ a constant damping parameter introduced to avoid strong oscillations during the shift evolution. This gauge condition 
not only maintains the well posedness of BSSN and CCZ4 formalisms, but also mimics the minimal distortion condition Eq.(\ref{betaellip}), trying then to minimize the stretching of the spatial coordinates. For numerical simulations involving black holes and neutron stars, standard values are $g(\alpha)=3/4$ and $\eta \approx 2/M$, being $M$ the mass of the compact object. Notice that in most of the literature the evolution of the shift is written in terms of an auxiliary field $B^i$, which however does not seem necessary for most of the relevant numerical scenarios~\cite{2006PhRvD..73l4011V}.

\section{Numerical methods}

In the same way that any reasonable physical model must be described by a well-posed PDE system, any numerical solution must satisfy the following three conditions: (i) \textit{consistency}, meaning that the discrete derivative operators reduce to the continuum ones as the resolution (i.e., the amount of discrete points sampling the continuum domain) increases; (ii) \textit{stability}, such that the numerical solution is bounded and depends continuously on the initial data, and (iii) \textit{convergence}, that is, the numerical solution tends to the continuum one as resolution increases.

Fortunately, it is not necessary to prove these three conditions, since Lax-Richtmyer equivalence theorem states that the numerical approximation of well-posed problems is convergent if and only if the scheme is stable and consistent~\cite{doi:10.1002/cpa.3160090206}. Since consistency can be obtained quite trivially, the relevant question here is how to discretize the equations such that the well-posedness at the continuum problem translate into stability at the discrete one. Let us consider the following generic set of hyperbolic PDE at the continuum,
\begin{eqnarray}
\partial_t u & = &  \mathcal{P} (u, \partial u)
\label{continuum_problem}
\end{eqnarray}  
where $\mathcal{P}$ is the evolution operator, which can depend on arbitrary spatial derivatives of $u$. A popular technique to  discretize  this continuum problem is by using the {\em Method of Lines} (MoL), which decouples the treatment of space and time coordinates~\cite{BA1279934X}. In the first step, only the spatial dimensions are discretized, while leaving the time continuous. This semi-discrete problem consist on a set of ordinary differential equations for $U_i(t) = u(t, x_i)$, one for each discrete spatial point $x_i$, separated by a mesh size $\Delta x$.  The semi-discrete equations can formally be written by substituting $\mathcal{P} \rightarrow P$, a discrete version of the evolution operator, written in terms of discrete derivative operators $D$.
In the second step, the fully discrete problem is obtained after discretizing in time, such that the fully discrete solution is given by $U^n_i=u(t^n, x_i)$ at each discrete time $t^n$, separated by a time-step $\Delta t$. The discrete equations can formally be written again by substituting $\partial_t$ by a discrete time integrator $D_t$.
As it is shown in~\cite{1995tpdm.book.....G}, the fully discrete problem preserves the stability of the semi-discrete problem if it is integrated with a locally-stable time integrator, as for instance any Runge-Kutta of at least $3^{rd}$ order. Thus, the problem is then reduced to ensure the stability of the semi-discrete problem by choosing a suitable space derivative discretization.

\subsection{Smooth solutions}

For sufficiently smooth solutions, the full procedure to ensure convergence of the numerical solution can be found in in~\cite{1995tpdm.book.....G} and summarized as follows: starting from a well-posed system at the continuum, apply the MoL, discretize in space with derivative operators satisfying certain conditions and then integrate with a Runge-Kutta of third order or higher. A problem at the continuum is well-posed if the solution 
satisfies an energy estimate which bounds some norm of the solution at some fixed time. A tool that is used in the derivation of such energy estimates is the integration by parts rule.
Analogously, a semi-discrete problem can be shown to be stable if the discrete difference operators $D$ satisfies the summation by parts rule, which is the discrete version of the integration by parts (see~\cite{2004CQGra..21.5735C} and references within for early works introducing these techniques in Numerical Relativity). 
For non-linear equations it is usually necessary to remove the high-frequency (unphysical) modes not accurately represented in the grid, which can grow continuously in time at any fixed resolution. The easiest way to damp these modes is by adding a filtering operator $Q_d U$ to the right-hand-side of the semi-discrete equations, like for instance the Kreiss-Oliger dissipation operator~\cite{574648}. This operator vanishes at infinite resolution, such that the semi-discrete problem is still consistent, and it is designed not to spoil the accuracy of the numerical scheme. Notice that not only Einstein equations, but any hyperbolic system of non-linear PDEs without the presence of either shocks or discontinuities, can be solved with these methods.

\subsection{Non-smooth solutions}

Although it is not the case with Einstein evolution systems, if the equations are genuinely non-linear like in fluid dynamics, discontinuities and shocks (i.e., a region with a crossing of the characteristics of the system) might appear even from smooth initial data. Discrete operators based on Taylor expansions, assuming smoothness of the solution, are going to fail near these regions and will produce artificial oscillations leading to unphysical solutions. Therefore, any spatial discretization able to handle shocks needs to take advantage of the integrated or weak-form of the equations~\cite{LeVeque1990}. Let us consider a system of non-linear PDEs, like the conservation of energy and momentum given by  Eqs.~(\ref{fluid31a},\ref{fluid31b}), which can be written in the following balance law form~\cite{2000PhRvD..61d4011F} 
\begin{eqnarray}
\partial_t u + \partial_k F^k(u) = S(u)
\label{balancelaw}
\end{eqnarray}  
where the fluxes $F^k(u)$ and the sources $S(u)$ depend on the fields but not on their derivatives. There are two popular different schemes to discretize these equations, based either on finite volumes or on finite differences~\cite{shu98}. 
The starting point of the finite-volume approach is the integral of the previous balance law equation in a spatial volume element $dV$,
\begin{eqnarray}
\partial_t {\bar u} + \oint F^k dS_k = {\bar S}
\label{balancelaw_weak}
\end{eqnarray}  
where ${\bar u}$ and ${\bar S}$ are the volume integrals of the corresponding quantity in the cell and we have used Gauss theorem to convert the volume integral of the fluxes into a a surface one, being $dS_k$ the surface element. 
This weak form can be easily discretized with a conservative scheme, namely
\begin{eqnarray}
\partial_t {\bar U}_i = - \frac{1}{\Delta x} \big[ F_{i+1/2} - F_{i-1/2} \big] + {\bar S}_i
\label{balancelaw_weak_discrete}
\end{eqnarray}  

\begin{figure}[h]
	\centering
	\includegraphics[width=7cm]{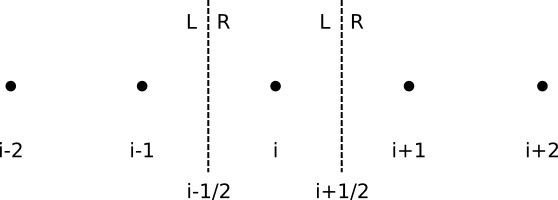}
	\caption{{\em{The computational uniform grid $x_i$}}. The left (L) and right (R) states reconstructed at the interfaces $x_{i\pm 1/2}$ are required to calculate the numerical flux $F_{i \pm 1/2}$. }
	\label{figure:weno}
\end{figure}

The problem is then reduced to compute (i) the solution at the grid points $U_i$ from the volume averages ${\bar U_i}$, and (ii) the numerical flux at the interfaces $F_{i\pm 1/2}$. These two steps must be performed in such a way that the semi-discrete solution is Total Variation Diminishing(TVD), or at least Total Variation Bounded(TVB), meaning essentially that no new extremes are allowed in the solution, which prevents the appearance of artificial oscillations. Notice that these conditions are more restrictive than stability, where the solution can still grow under certain tolerant bounds. The procedure to construct a shock-capturing scheme is the following. First, one needs to reconstruct the fields at the interfaces $x_{i\pm 1/2}$ using information either from the right (R) or from the left (L) to the interface (see Figure~\ref{figure:weno}), namely $(u^R_{i\pm 1/2}, u^L_{i\pm 1/2})$. Commonly used high-order reconstructions, preserving the monotonicity of the solution to prevent spurious oscillations, are for example the Weighted-Essentially-Non-Oscillatory (WENO) reconstructions~\cite{jiang96,shu98} and MP5~\cite{suresh97}.
Then, a suitable flux-formula is required to solve, at least approximately, the jump on the fields at each interface (i.e., Riemann problem), by combining information from the right and from the left, namely $F_{i\pm 1/2} = F (u^R_{i\pm 1/2}, u^L_{i\pm 1/2})$. This flux-formula usually requires information on the characteristic structure of the system (i.e., eigenvectors and eigenvalues). 
This approach has been the most commonly employed in binary neutron star simulations, see for instance~\cite{2002PThPh.107..265S,2008PhRvD..77b4006A,2008PhRvD..78b4012L,2008PhRvD..78h4033B,2008PhRvD..78f4054Y,2014CQGra..31a5005M}.

Higher-order schemes are relatively easy to achieve with the finite-difference approach, providing an efficient approach to high-order shock-capturing methods~\cite{10.1016/0021-9991(88)90177-5}. However, high-order finite-difference numerical schemes applied to the magneto-hydrodynamics (MHD) equations have not been as robust as those based on finite-volume. Nowadays that is not a great inconvenient, and the possibility to achieve high order accuracy is leading to more efforts on implementing these methods on computational MHD codes~\cite{2014CQGra..31g5012R,2016PhRvD..94f4062B}. Although the derivation is different, the conservative scheme given by Eq.(\ref{balancelaw_weak_discrete}) is still valid,
where now ${\bar U}$ means just $U_i$, the value of the field in the grid point. Again, the problem is reduced to compute a suitable numerical flux at the interfaces such that solution is essentially non-oscillatory and preserves, or at least bounds, the Total Variation. The procedure starts by performing a Lax-Friedrichs splitting, where it is introduced the following combination of fluxes and fields $F^{\pm}_i = 1/2 (F_i \pm \lambda U_i)$, being $\lambda$ the maximum eigenvalue in the neighborhood of the point. These combinations are interpolated at the interfaces by using a monotonic reconstruction, like the high-order ones discussed before. The flux at the left of the interface $F^L_{i+1/2}$ is reconstructed using the values $\{F^+\}$, while that the flux at the right $F^R_{i+1/2}$ is reconstructed using the values $\{F^-\}$. The final numerical-flux is obtained just as $F_{i +1/2} = F^R_{i + 1/2} + F^L_{i + 1/2}$.
At the lowest order reconstruction, $F^L_{i+1/2}=F^+_i$ and $F^R_{i+1/2}=F^-_{i+1}$, so that the final numerical-flux reduces to the popular and robust Local-Lax-Friedrichs flux~\cite{LeVeque1990}.

Finally, notice that efforts considering other techniques to solve self-gravitation neutron stars, like the discontinuous Galerkin  methods~\cite{2018PhRvD..98d4041H}, are underway and might be an interesting option in the near future.

\section*{Acknowledgments}

It is my pleasure to thank Carles Bona, Miguel Bezares, Luis Lehner and Joan Mass\'o for their critical reading and helpful comments  on the manuscript.
I acknowledge support from the Spanish Ministry of Economy and Competitiveness grant AYA2016-80289-P (AEI/FEDER, UE).


\bibliographystyle{utphys}
\bibliography{review}

\end{document}

%% file: aas_macros.tex
\def\jnl@style{\it}
\def\aaref@jnl#1{{\jnl@style#1}}

\def\aaref@jnl#1{{\jnl@style#1}}

\def\aj{\aaref@jnl{AJ}}                   
\def\araa{\aaref@jnl{ARA\&A}}             
\def\apj{\aaref@jnl{ApJ}}                 
\def\apjl{\aaref@jnl{ApJ}}                
\def\apjs{\aaref@jnl{ApJS}}               
\def\ao{\aaref@jnl{Appl.~Opt.}}           
\def\apss{\aaref@jnl{Ap\&SS}}             
\def\aap{\aaref@jnl{A\&A}}                
\def\aapr{\aaref@jnl{A\&A~Rev.}}          
\def\aaps{\aaref@jnl{A\&AS}}              
\def\azh{\aaref@jnl{AZh}}                 
\def\baas{\aaref@jnl{BAAS}}               
\def\jrasc{\aaref@jnl{JRASC}}             
\def\memras{\aaref@jnl{MmRAS}}            
\def\mnras{\aaref@jnl{MNRAS}}             
\def\pra{\aaref@jnl{Phys.~Rev.~A}}        
\def\prb{\aaref@jnl{Phys.~Rev.~B}}        
\def\prc{\aaref@jnl{Phys.~Rev.~C}}        
\def\prd{\aaref@jnl{Phys.~Rev.~D}}        
\def\pre{\aaref@jnl{Phys.~Rev.~E}}        
\def\prl{\aaref@jnl{Phys.~Rev.~Lett.}}    
\def\pasp{\aaref@jnl{PASP}}               
\def\pasj{\aaref@jnl{PASJ}}               
\def\qjras{\aaref@jnl{QJRAS}}             
\def\skytel{\aaref@jnl{S\&T}}             
\def\solphys{\aaref@jnl{Sol.~Phys.}}      
\def\sovast{\aaref@jnl{Soviet~Ast.}}      
\def\ssr{\aaref@jnl{Space~Sci.~Rev.}}     
\def\zap{\aaref@jnl{ZAp}}                 
\def\nat{\aaref@jnl{Nature}}              
\def\iaucirc{\aaref@jnl{IAU~Circ.}}       
\def\aplett{\aaref@jnl{Astrophys.~Lett.}} 
\def\apspr{\aaref@jnl{Astrophys.~Space~Phys.~Res.}}
\def\bain{\aaref@jnl{Bull.~Astron.~Inst.~Netherlands}} 
\def\fcp{\aaref@jnl{Fund.~Cosmic~Phys.}}  
\def\gca{\aaref@jnl{Geochim.~Cosmochim.~Acta}}   
\def\grl{\aaref@jnl{Geophys.~Res.~Lett.}} 
\def\jcp{\aaref@jnl{J.~Chem.~Phys.}}      
\def\jgr{\aaref@jnl{J.~Geophys.~Res.}}    
\def\jqsrt{\aaref@jnl{J.~Quant.~Spec.~Radiat.~Transf.}}
\def\memsai{\aaref@jnl{Mem.~Soc.~Astron.~Italiana}}
\def\nphysa{\aaref@jnl{Nucl.~Phys.~A}}   
\def\physrep{\aaref@jnl{Phys.~Rep.}}   
\def\physscr{\aaref@jnl{Phys.~Scr}}   
\def\planss{\aaref@jnl{Planet.~Space~Sci.}}   
\def\procspie{\aaref@jnl{Proc.~SPIE}}   